\documentclass[prb,showpacs,twocolumn,superscriptaddress]{revtex4}
\usepackage{hyperref,graphicx,dcolumn}
\usepackage{amsfonts,amsmath,amssymb,bm}
\usepackage{color}
\usepackage{multirow}
\usepackage{braket}
\usepackage{lipsum}

\newcommand{\myFig}[6]{ %
\begin{figure}[htb] 
\begin{center} 
\includegraphics[width=#1\columnwidth,height=#2\columnwidth,clip=true,keepaspectratio=#3]{#4}
\caption{#5} \vspace{-0.5cm} \label{#6} 
\end{center} \end{figure}} 

\newcommand{\vcr}[1]{\boldsymbol{\mathrm{#1}}}

\newcommand{\scal}[2]{\left\langle #1 \vert #2 \right\rangle}

\newcommand{\Tr}[1]{ \mathrm{Tr} \left[ #1 \right]}
\newcommand{\SqB}[1]{ \left[ #1 \right]}
\newcommand{\RnB}[1]{ \left( #1 \right)}
\newcommand{\CrB}[1]{ \left\{ #1 \right\}}

\begin{document}
\title{First-principles spin-transfer torque in CuMnAs$\vert$GaP$\vert$CuMnAs junctions}
\author{Maria Stamenova}\email[Contact email address: ]{stamenom@tcd.ie}
\affiliation{School of Physics, AMBER and CRANN Institute, Trinity College, Dublin 2, Ireland} 
\author{Razie Mohebbi} 
\affiliation{Vali-e-Asr University, Rafsanjan, Iran} 
\author{Jamileh Seyedyazdi}
\affiliation{Vali-e-Asr University, Rafsanjan, Iran}
\author{Ivan Rungger}
\affiliation{National Physical Laboratory, Teddington, TW11 0LW, United Kingdom}
\author{Stefano Sanvito} 
\affiliation{School of Physics, AMBER and CRANN Institute, Trinity College, Dublin 2, Ireland} 

\begin{abstract}
We demonstrate that an all-antiferromagnetic tunnel junction with current perpendicular to the plane geometry can be 
used as an efficient spintronics device with potential high frequency operation. By using state of the art density functional
theory combined with quantum transport, we show that the N\'eel vector of the electrodes can be manipulated by 
spin-transfer torque. This is staggered over the two different magnetic sublattices and can generate dynamics and switching. At 
the same time the different magnetization states of the junction can be read by standard tunnelling magnetoresistance. 
Calculations are performed for CuMnAs$\vert$GaP$\vert$CuMnAs junctions with different surface terminations between 
the anti-ferromagnetic CuMnAs electrodes and the insulating GaP spacer. In particular we find that the torque remains 
staggered regardless of the termination, while the magnetoresistance depends on the microscopic details of the interface.
\end{abstract}


\maketitle

%\section{Introduction}

Antiferromagnetic (AF) materials are magnetically ordered compounds where two or more spin sublattices 
compensate each other resulting in a vanishing macroscopic magnetization. As a consequence an 
antiferromagnet does not produce stray field and closely separated AF nanostructures are not magneto-statically 
coupled. In addition the typical time scale for the dynamics of the antiferromagnetic order parameter, the N\'eel 
vector, is set by the AF resonance frequency, which is typically much larger than that of a ferromagnet and may 
approach the THz limit~\cite{CoeyBook}. It is then not surprising that antiferromagnets have recently received 
considerable attention as a materials platform for magnetic data storage, logic and high-frequency 
applications~\cite{RevAFSpin}. A significant limitation of the AF materials class is the fact that most antiferromagnets 
are insulators, while often spintronics devices require driving currents through the structure. 

Recently metallic CuMnAs has been proposed as a good candidate for AF spintronics applications~\cite{CMA_Nat}. 
Tetragonal CuMnAs is antiferromagnetic at room temperature and can be grown epitaxially on GaP. Furthermore, it has 
been shown that one can manipulate the N\'eel vector of CuMnAs thin films by electric current pulses \cite{CMA_Science}.
This is explained as the result of atomically-staggered spin-orbit torques (SOTs) \footnote{These are also referred to as N\'eel-order spin-orbit torque fields \cite{Zelezny}.}, which accompany the current flow in antiferromagnets where the global inversion symmetry is broken due to the presence of two spin sublattices forming inversion 
partners~\cite{Zelezny}. The N\'eel temperature of CuMnAs is reported~\cite{CMA_SciRep} to be $(480 \pm 5)$~K, while 
the lattice parameters of bulk tetragonal CuMnAs are $a=b=3.820$~\AA~and $c=6.318$~\AA. According to density
functional theory (DFT) calculations CuMnAs in its AF ground state is metallic but its very low density of states at the 
Fermi level, $E_\mathrm{F}$, resembles that of a semi-metal or a narrow-band semiconductor~\cite{CMA_Nat}. Here 
we investigate whether such unique AF metal can be used in standard magnetic tunnel junctions (MTJ) and demonstrate 
that these can be written by spin-transfer torques (STTs) and read by standard tunnel magnetoresistance (TMR).

\myFig{1}{1}{true}{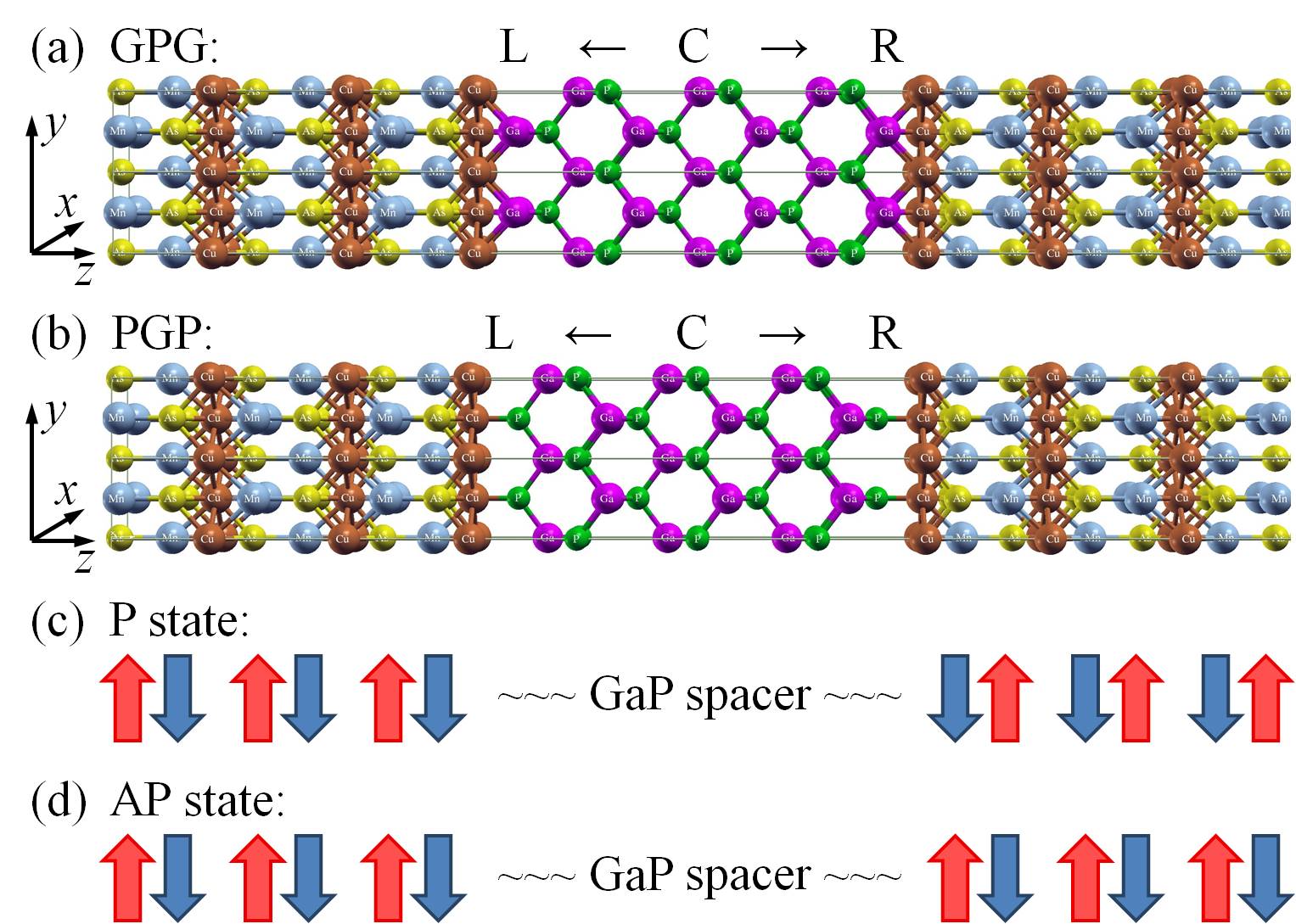}{(Color online) Atomistic diagrams of the two CuMnAs$\vert$GaP$\vert$CuMnAs junctions 
studied here: (a) Ga-terminated structure (labelled as GPG) and (b) P-terminated structure (PGP). The position of 
the central atomic layer is marked by ``C'' and the labels ``L'' and ``R'' stand for the left- and right-hand-side half of 
the junction. Panels (c) and (d) show cartoons of the two possible orientations of the Mn spins in the junction. 
We define a parallel (P) and anti-parallel (AP) state of the junction from the orientation of the two innermost Mn
planes forming the interface with GaP. The arrows indicate the direction of the spin of the Mn atoms in a corresponding 
layer. Hence, in the AP case the right-hand-side lead (R) order parameter is rotated by 180$^\mathrm{o}$ with respect 
to the P state. Coordinate system is chosen such that the $z$-axis is along the transport direction, while the N\'eel vectors of the leads are rotated in the $(x-y)$ plane.}{fig01}

We consider the two CuMnAs$\vert$GaP$\vert$CuMnAs stacks depicted in Fig.~\ref{fig01}. These are built along the 
CuMnAs [001] direction with the GaP spacer taking a zincblende structure. In both cases CuMnAs is terminated at the Cu
plane and the two electrodes are atomically mirror-symmetric with respect to the central $(x-y)$ plane in the junction. 
The spacer can then be terminated either at the Ga-plane (GPG junction) or 
at the P one (PGP). In both cases the total number of atoms in the junction supercell is the same -- 49 atoms. Note that the 
GPG structure does not only differ from the PGP by the GaP termination, but also because P is on top of Cu, 
while Ga is placed at the hollow site. Further details on the geometry of the junctions is included in the supporting information (SI).      

The local spin-density approximation (LSDA) is able to reproduce the experimentally observed AF structure of 
CuMnAs~\cite{CMA_Nat} in its tetragonal phase for the experimental lattice parameters. Both the GPG and the PGP 
junction consist of 3 CuMnAs unit cells at each side of a 3 unit-cell-long GaP spacer region. In addition there is an 
extra Ga/P layer at one of the interfaces in order to make the junctions more symmetric. The whole junctions are relaxed in the $z$-direction within LSDA using periodic boundary conditions in all directions with the {\sc Siesta}~\cite{siesta} code. After the relaxation the $z$-coordinates of all atoms in both junctions are additionally symmetrized with respect to the junction center. In this way we suppress additional geometry-driven asymmetry effects to the transport properties. In the final geometries, the atoms from the leads map ideally 
onto the same species upon reflection in $(x-y)$ plan through the central atom, and the same is true also for the $z$-coordinates (only!)
of the spacer atoms. The $x$ and $y$ coordinates of the spacer atoms adjacent to the interfaces are different. In the 
GPG junction the Cu-Ga distance along the $z$-axis is approximately 1.75 \AA~for both interfaces, while in the PGP 
junction the Cu-P distance is 2.16 \AA~for both interfaces. The total length of the junction is 54.5~\AA\ for the GPG and 
54.9~\AA\ for the PGP. 

%\section{Theory}

Electron transport through the junctions is described by the Keldysh non-equilibrium Green function (NEGF) method as 
implemented in the {\sc Smeagol} code \cite{smeagol1,smeagol2,smeagol3}. We consider a steady-state formulation of 
the STT analogous to that described in Ref.~[\onlinecite{HaneyCoCu}], that we briefly recall here. In the NEGF formalism 
the semi-infinite periodic electrodes to the left (L) and to the right (R) of the central scattering region (C) are described by energy
dependent complex self-energies, $\Sigma_\mathrm{L,R} (E)$, while the Green's function (GF) of the open-boundary system 
reads
\begin{equation}\label{green}
{G}(E)=\lim_{\eta\rightarrow 0} \left[\left(E+i\eta\right)-H_{\mathrm{C}}(E)-{\Sigma}_\mathrm{L}(E) -
{\Sigma}_\mathrm{R}\right(E)]^{-1}\:,
\end{equation}
where $\Sigma_\mathrm{L}=H_\mathrm{LC}^\dagger{g}_{\mathrm{L}}H_\mathrm{LC}$, 
$\Sigma_\mathrm{R}=H_\mathrm{RC}{g}_{\mathrm{R}}H_\mathrm{RC}^\dagger$, ${g}_{\mathrm{R,L}}(E)$ is the 
surface GFs of the leads, while $H_\mathrm{\alpha C}$ ($\alpha=$~R, L) are the Hamiltonian matrix blocks describing 
the interaction between the leads and the central region. Since ${G}(E)$ is a finite non-Hermitian matrix the number of 
electrons in the scattering region is not conserved and the associated density matrix
\begin{equation}
\label{den_mat}
\rho=\frac{1}{\pi}\int_{-\infty}^{\infty} dE \left[\rho_\mathrm{L}(E)\eta^\mathrm{L}_\mathrm{F}(E) + \rho_\mathrm{R}(E)
\eta^\mathrm{R}_\mathrm{F}(E) \right] \;.
\end{equation} 
must be evaluated self-consistently. In Eq.~(\ref{den_mat}) $\rho_\alpha(E)=G (E)\Gamma_\alpha (E) G^{\dagger}(E)$ 
are the partial density-of-states operators for electrons originating from each lead, $\Gamma(E)=i [\Sigma(E)-\Sigma^{\dagger}(E)]/2$ 
is the non-Hermitian part of the self energy and $\eta^\alpha_\mathrm{F}(E)=\eta_\mathrm{F}(E-\mu_\alpha)$ are the corresponding 
Fermi distributions for the electron reservoirs with chemical potential $\mu_\alpha$. The transmission is defined as
\begin{equation}
t^\sigma (E)=\Tr{\Gamma_L (E) G^{\dagger} (E) \Gamma_R (E) G (E)  }^{\sigma\sigma}
\end{equation}
for each spin channel $\sigma=\CrB{\uparrow,\downarrow}$.

We extend the STT definition of Ref.~[\onlinecite{HaneyCoCu}] to a spin-DFT implementation based on non-orthogonal local-orbital
basis sets. The spin-polarized part of the Kohn-Sham (KS) Hamiltonian, which originates from the exchange-correlation potential, 
can be isolated as $\vcr{\Delta}_{ij}=\sum_{\alpha,\beta}H^{\alpha\beta}_{ij}\vcr{\sigma}^{\beta\alpha}$, where Latin letters ($i,j$) 
index matrix elements with respect to the localized basis, Greek letters are the spin indices and $\vcr{\sigma}$ is the vector of 
Pauli matrices. Analogously the spin part of the density matrix is 
$\vcr{s}_{ij}=\sum_{\alpha,\beta}\rho^{\alpha\beta}_{ij}\vcr{\sigma}^{\beta\alpha}$ and the total spin in the system is 
$\vcr{S}_\mathrm{tot}=\Tr{\rho\vcr{\sigma}\Omega}=\sum_{i,j} \vcr{s}_{ij}\Omega_{ji}$, where $\Omega_{ij}=\scal{\phi_i}{\phi_j}$ is the basis 
overlap matrix. The M\"ulliken-partitioned local atomic spin at a site $a$ is thus defined as 
$\vcr{S}_a=\Re\SqB{\sum_{i\in a}\sum_j \vcr{s}_{ij} \Omega_{ji} }$ and the local spin torque for a closed system is
\begin{equation}
\vcr{T}_a= \frac{\hbar}{2} \dot{\vcr{S}}_a = \frac{\hbar}{2}\sum_{i \in a}\sum_j \mathrm{Tr}\SqB { \dot{\rho}_{ij}\Omega_{ji}\vcr{\sigma} }\, .
\end{equation} 
thus giving
\begin{equation}
\vcr{T}_a=  \frac{1}{2}\sum_{i \in a}\sum_j \SqB{ \Im\RnB{H^0_{ij} \vcr{s}_{ij} - \rho^0_{ij}\vcr{\Delta}_{ji}} + \Re\RnB{\vcr{\Delta}_{ij}\times\vcr{s}_{ji}} }   \, ,
\end{equation}
where we have used the Liouville equation for the density matrix in the non-orthogonal basis $\dot{\rho}=\frac{1}{i\hbar}\RnB{\Omega^{-1}H\rho - \rho H \Omega^{-1}}$ and the superscript ``0'' indicates the spin-independent part of the operator, i.e. $\hat{H}^0=\hat{H}-\frac{1}{2}\hat{\vcr{\Delta}}\cdot\hat{\vcr{\sigma}}$. Note that all spin-dependent interaction is carried by the exchange-correlation field $\hat{\vcr{\Delta}}$ and the spin-orbit interaction is neglected in our description. 

As in Ref.~[\onlinecite{HaneyCoCu}] we can extend this result to an open-boundary system and define the STT as the 
\begin{equation}
\vcr{T}_i =\frac{1}{2} \sum_{i \in a}\sum_j \vcr{\Delta}_{ij} \times \vcr{\sigma}_{tr,ji}
\end{equation} 
where $\vcr{\sigma}_{\mathrm{tr},i}$ is the non-equilibrium (transport) spin due to some inherent spin non-collinearity in the 
open-boundary system. Typically in a steady state current-carrying condition $\vcr{\sigma}_\mathrm{tr}$ is taken as 
$\Tr{\RnB{\rho_V-\rho_0}\vcr{\sigma}}$ where $\rho_V$ and $\rho_0$ are the non-equilibrium (finite bias) and the equilibrium 
(zero bias) density matrices, respectively. If we assume a small perturbation $\delta V$ of the chemical potentials of the leads 
such that  $\mu_\mathrm{L} \rightarrow \mu_\mathrm{L} + \delta V/2$ while $\mu_\mathrm{R} \rightarrow \mu_\mathrm{R}- \delta V/2$, 
the non-equilibrium charge density induced by $\delta V$ is
\begin{equation}
\rho_\mathrm{tr} (\delta V) \approx \Tr{\frac{\partial \rho (V)}{\partial V}}_{V=0} \, \delta V\:,
\end{equation}    
and the corresponding linear-response spin density is then
\begin{equation}
\vcr{\sigma}_\mathrm{tr} (\delta V) \approx \Tr{\frac{\partial \rho (V)}{\partial V} \vcr{\sigma}}_{V=0} \, \delta V \, .
\end{equation} 
Based on this derivation we can evaluate the so-called spin-transfer {\it torkance} (STTk) as
\begin{equation}
\vcr{\tau}_a =  \vcr{T}_a/\delta V = \frac{1}{2}\sum_{i \in a} \sum_j\vcr{\Delta}_{ij}\times \Tr{\frac{\partial \rho_{ji} (V)}{\partial V} \vcr{\sigma}}_{V=0}  \, .
\end{equation}
By assuming that $G(E)$ and $\Gamma(E)$ are slowly varying functions around the Fermi level the derivate above takes the form
\begin{equation}
\left. \frac{\partial \rho (V)}{\partial V} \right\vert_{V=0} = \frac{1}{4\pi} G(E_F) \SqB{\Gamma_L(E_F) - \Gamma_R(E_F)} G^\dagger (E_F) \,.
\end{equation}

%\section{Results}

\myFig{1}{1}{true}{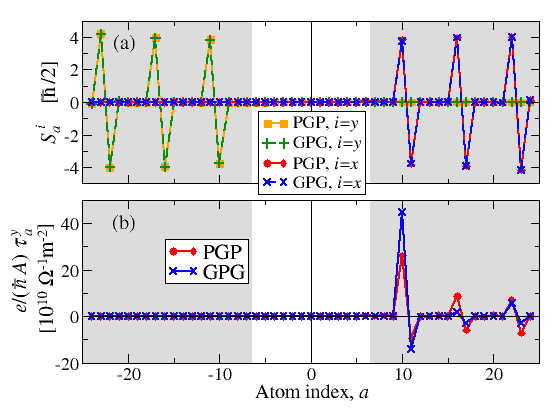}{(Color online ) Ground state observables atomically-projected along the $z$-direction in the junction for 
the 90$^\mathrm{o}$ alignment of the leads order parameter in the PGP and the GPG geometries: (a) the $x$ and $y$-components of the 
local spin moments and (b) the spin-transfer torkance per unit area (for both junctions $A=14.59$ \AA$^2$). }{fig02}

We evaluate the atomically-resolved SSTk in the right-hand side lead for the largest misalignment of 90$^\mathrm{o}$ between the 
leads order parameters, i.e for the Mn spins in the left-hand side lead (LL) oriented along the $y$-axis, while the spins in the right-hand side 
lead (RL) are rotated by 90$^\mathrm{o}$ about the $z$-axis starting from the P state. In this configuration the order parameter in RL is now 
along $x$ [Fig.~\ref{fig02} (a)]. The local exchange field $\vcr{\Delta}_a$ within LSDA shows a very similar staggered distribution (not shown here) and the resulting STTk in the right-hand side lead is also staggered.
%There are very little differences in the magnitude of the onsite Mn spins and exchange fields between the two junction types. We find, however, a more pronounced difference in the out-of-plane transport spin distributions, $\vcr{\sigma}_{\mathrm{tr},a}=\sum_{i \in a} \vcr{\sigma}_{\mathrm{tr},ii}$, which both show oscillations in the spacer and into the leads. For both junctions there are pronounced peaks at the innermost Mn sites and in the right-hand side lead the pairs of Mn atoms have the same sign $\sigma^z_{\mathrm{tr},a}$. This, together with the sign-alternating $\Delta_a^x$, gives rise to staggered in-plane STTk for both junctions.
It is maximal at the very first Mn layer and then decreases in amplitude in the following two Mn bi-layers. Despite the limiting size of the MTJ stack considered here, this provides strong indications of decay of the staggered in-plane STTk inside the electrode. This behavior of the SSTk is similar to what has been found by first principles calculations in AF spin-valves \cite{HaneyAF} and attributed to the interference of the multiple open channels with different $k_z$ at the Fermi level. It had been previously demonstrated that in the limit of a single open channel in model AF junctions the non-equilibrium spin density ($\sigma^{z}_{\mathrm{tr},a}$) is uniformly distributed in the lead \cite{AFmodel}, but this picture breaks down in the case of multi-channel conductance. 
%Note that there is a level of inaccuracy regarding transport properties like the STTk at the very boundaries (last unit cell of the lead buffer) of the finite junction which is inherent to the application of the boundary conditions and the finite-range self-energies. 
Most importantly, the observed in both junctions staggered property of the STTk, which is crucial for the manipulation of the AF order parameter by currents, appears to be a robust property of the AF junctions and it is weakly affected by the chemistry of the interfaces. The magnitude of the STTk we find is comparable (for similarly thick barriers) to those obtained by similar first-principles calculations in Fe$\vert$MgO$\vert$Fe MTJs \cite{Stiles08}. 
  
\myFig{1}{1}{true}{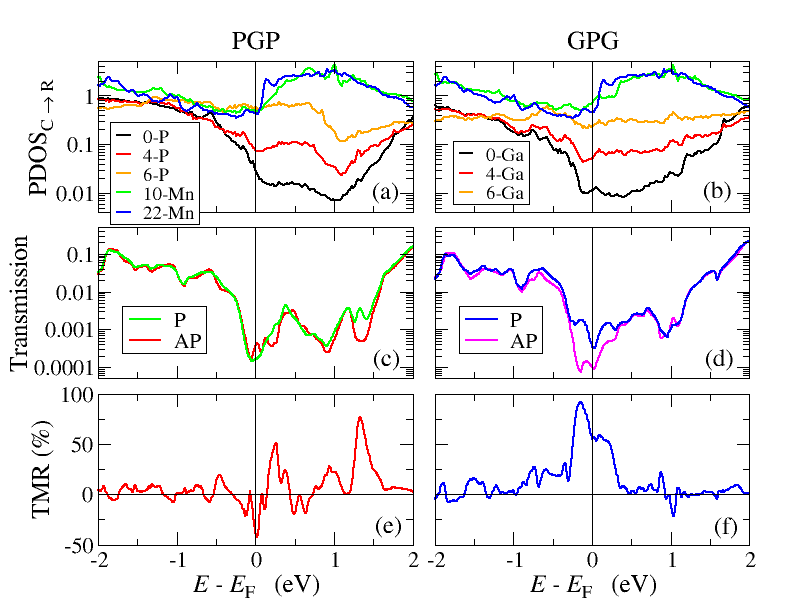}{(Color online) Electronic and transport properties of the two junctions in the P state: (a,b) the PDOS of 
atoms in the (C$\rightarrow$R)-half of the junction. The atom indexes are the same 
as in Fig.~\ref{fig02}. Note that the black curves corresponding to the atom (P or Ga) in the very center of the junctions. Panels (c,d) show 
the spectrum of the total transmission in both the P and AP state, while in panels (e,f) -- the TMR defined as 
$\mathrm{TMR}= \RnB{T_\mathrm{P}-T_\mathrm{AP}}/\RnB{T_\mathrm{P}+T_\mathrm{AP}}$. }{fig03}

Let us now investigate in more details for electronic and transport properties of the junction. The atomically-projected density of 
states (PDOS) in Fig.~\ref{fig03}(a,b) shows that the P interface layer is almost metalized, and there is a progressive reduction of 
the P PDOS when moving away from the interface for energies between $E_\mathrm{F}$ and $E_\mathrm{F}+2$ eV. This range
corresponds to the energy gap in the middle of the GaP spacer. The Fermi level is within the GaP gap, close to the valence 
band top, so that transport in via tunnelling. As such, the transmission coefficient, $t=t^\uparrow+t^\downarrow$, is reduced in 
the energy range of the gap [Fig.~\ref{fig03}(c,d)], and the main features are the same for GPG and PGP, as well as for the P and 
AP configurations. Importantly, for the GPG junction there is a significant energy interval $\pm$0.3 eV around $E_\mathrm{F}$, 
where the P configuration has significantly higher $t$ than the AP one, resulting in the large TMR of 50-100\% [Fig.~\ref{fig03}(c,d)]. 
In contrast, in the PGP junction the TMR does not display such an extended plateau but rather shows an oscillating energy dependance, 
and exactly at $E_\mathrm{F}$ there is a TMR sign reversal when compared to the GPG case. This is followed by an energy range 
above $E_\mathrm{F}$ where TMR$>$0 and then another region with significant TMR corresponding to the bottom of the GaP 
conduction band. Such oscillating character of the TMR in the PGP junction suggests that the actual value may be sensitive to 
the exact details of the atomic interfaces, while for the GPG case it can be expected to be robust. 

In order to obtain further insights we compare the transmissions for the two junctions, $t_\mathrm{\bf{k}}$, at the Fermi level as a function 
of the transverse $\vcr{k}_\perp=(k_x,k_y)$ vector resolved over the entire first Brillouin zone (BZ) (see Fig.~\ref{fig04}). We note that in both
cases there are vast regions of the Brillouin zone, where $t_\mathrm{\bf{k}}=0$ (blue areas in the transmission panels). This is a direct
consequence of the low DOS in the CuMnAs electrodes around $E_\mathrm{F}$. In fact in Fig.~\ref{fig04}(a) one can immediately observe
that the regions of $k$-space with $t_\mathrm{\bf{k}}=0$ essentially corresponds to regions where there are no open scattering 
channels at the Fermi level. Interestingly and at variance with the case of Fe/MgO junctions, such property interests also the BZ center.
\myFig{1}{1}{true}{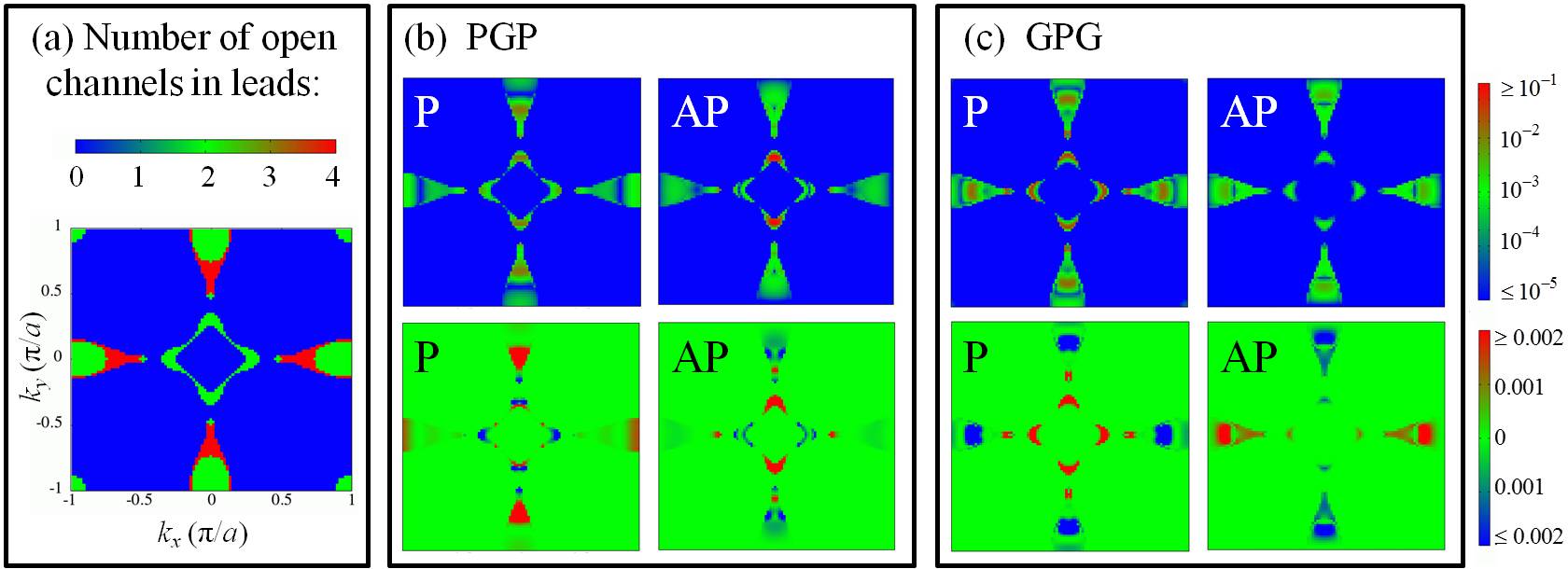}{(Color online) 2D reciprocal space, $(k_x,k_y)$, contour plots in the 2D Brillouin zone orthogonal to the 
transport direction for various transport quantities. Panel (a) depicts total number of open channels in the CuMnAs leads, (b,c) show 
the Fermi-level transmission $t(k_x,k_y)$ (top panels) and the spin-polarized transmission $\Tr{\hat{t}\hat{\sigma}^z}$ (bottom panels) 
in the P and AP state and for the PGP and GPG junction, respectively.}{fig04}

It is also interesting to note that the spin-polarized transmission, defined as $\Tr{\hat{t}\hat{\sigma}^z}$, shows that for different pockets in 
the BZ the spin of the current-carrying electrons is opposite. For example, for the GPG junction and P configuration the electrons have up-spin 
for the pockets close to the BZ center, while they have down-spin for the pockets at the BZ boundary. In contrast, for the AP configuration the 
pockets at  $k_y=0$ have up spins, while the ones at $k_x=0$ have down spins, so that the four-fold rotational symmetry in the $x-y$ plane is 
broken. This is due to the fact that the GPG structure breaks the symmetry in the $x-y$ plane, since the Ga-P bonds at the two 
interfaces are $90^\mathrm{o}$-rotated with respect to each other. The strong influence of such asymmetry on the transmission 
suggests a strong contribution of $d$-orbitals around $E_\mathrm{F}$.

In the supporting information (SI) we provide a detailed analysis of the $k$-dependent PDOS around $E_\mathrm{F}$, and we find evidence that, while for the GPG structure the transmission properties are mainly driven by bulk-like features, for the PGP one they are determined by interface states. The interface Mn layer has positive Fermi level spin polarization for the P alignment, while for the AP configuration the interface Mn has negative spin polarization, and only in the second layer the Mn moment is positively polarized. The effective tunnelling distance for each spin-configuration for the bulk-like states of GPG is therefore increased for the AP configuration compared to the P one, so that one expects a smaller conductance for the AP in this case.

\myFig{1}{1}{true}{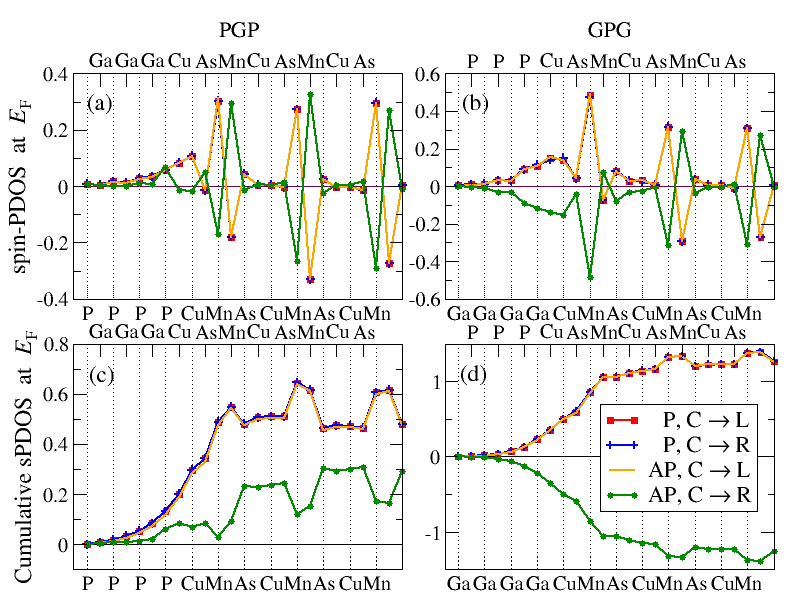}{(Color online) (a,b) The atomically-resolved spin-PDOS at the Fermi level as a function of atom position on the left-hand or 
right-hand side of the center of the junction for the two spin configurations (P and AP) and both PGP and GPG junctions. Panels (c,d) show the 
cumulative spin-PDOS, integrated from the center of the junction in all cases. Note that the legend for the symbols and line colors for all panels 
is in panel (d).}{fig05}

Finally, we look at the PDOS at the Fermi level and in particular the spin-PDOS, which we define as the PDOS difference for the spin-up and spin-down 
states~\footnote{The PDOS is defined as $N^\sigma (E)=\frac{1}{2\pi}\Tr{A^\sigma (E) \Omega}$, 
where $A^\sigma(E)=2\pi\sum_{n} \delta\RnB{E-E^\sigma_n}\psi_n^\sigma\psi_n^\sigma\dagger$ is the spectral function for each spin component 
$\sigma=\uparrow,\downarrow$ and $\psi_n^{\sigma}$ is the KS solutions of the open-boundary problem corresponding to eigenvalue $E^\sigma_n$.} 
as a function of the atomic position starting from the center of the junction (see Fig. \ref{fig05}). We find the staggered Fermi-level spin-PDOS at the Mn atoms which is in fact oppositely polarized to the on-site spin (condensate) at each Mn atom (see Fig. \ref{fig01}). 
Most importantly, there is a notable difference between the two junctions. For both the PGP and GPG stacks the spin-PDOS curves for the P configurations 
and for one of the AP are identical. However, the curve for the remaining AP configuration (C$\rightarrow$R) has a cumulative spin-PDOS with opposite sign
for the GPG junction and with the same sign for the PGP. Thus in the GPG stack the spin polarization induced inside the spacer around each interface carries the 
same sign as the Mn atom at the interface, while this is not the case for the PGP junction in the AP state. This suggests that the TMR in the GPG structure 
resembles the conventional TMR of ferromagnetic junctions as described by the Julli\`ere model~\cite{Julliere}. In fact, it can be seen in Fig. \ref{fig05}(d) 
that the PDOS for majority and minority spins swap as the non-compensated spin-polarization reverses in the right-hand side lead between the P and 
the AP state.     

In conclusion we have shown that stacks made entirely of anti-ferromagnets display staggered spin-transfer torques when the current flows in 
a perpendicular to the plane orientation. This fact is little affected by surface termination and strongly suggests that the antiferromagnetic
order parameter can be manipulated by currents. Furthermore, such junctions exhibit pronounced magnetoresistance, which is intrinsic 
of having asymmetric stacks. In particular here we have demonstrated that for CuMnAs/GaP structures a Ga-Cu termination enhances the
magnetoresistance. Our work thus demonstrates that all-antiferromagnetic junctions are both readable and writable with an electrical current,
and therefore are interesting candidates as high-frequency high-density memory elements. 

\subsection*{Acknowledgment}

This work is supported by Science Foundation Ireland (grant No. 14/IA/2624) and from the European Union's Horizon2020 research and innovation programme within the PETMEM project (grant No. 688282). We gratefully acknowledge the DJEI/DES/SFI/HEA Irish Centre for High-End Computing (ICHEC) for the provision of computational facilities. Some of the calculations were performed on the Parsons cluster maintained by the Trinity Centre for High Performance Computing. This cluster was funded through grants from Science Foundation Ireland.

\clearpage
\onecolumngrid

\section{Supplemental materials}
\subsection{Interface geometry}

We present in Fig. \ref{fig06} details of the geometry of the two interfaces, labeled as ``Left'' (or just L) and ``Right'' (or R) of the two junctions 
presented in Fig. \ref{fig01}. Note that the junctions have been rotated by $\pm 90^\mathrm{o}$ around the $x$-axis on Fig. \ref{fig01} but there 
has been no rotation around the $z$-axis. Hence the first Ga-P bonds after the Cu interface are in the $(x-z)$ plane for the L or in the $(y-z)$ plane 
for the R. The opposite is for the GPG junction. Although the exact composition of the CuMnAs|GaP interface cannot be unequivocally deduced from 
the TEM images contained in Ref. \onlinecite{CMA_Science}, the growth orientation of the tertragonal CuMnAs on zinc blende GaP has been described 
in Ref. \onlinecite{CMA_Nat}. This has been the motivation for the interface reconstructions we use, i.e. P on top of Cu, while Ga on hollow site. 
\myFig{1}{1}{true}{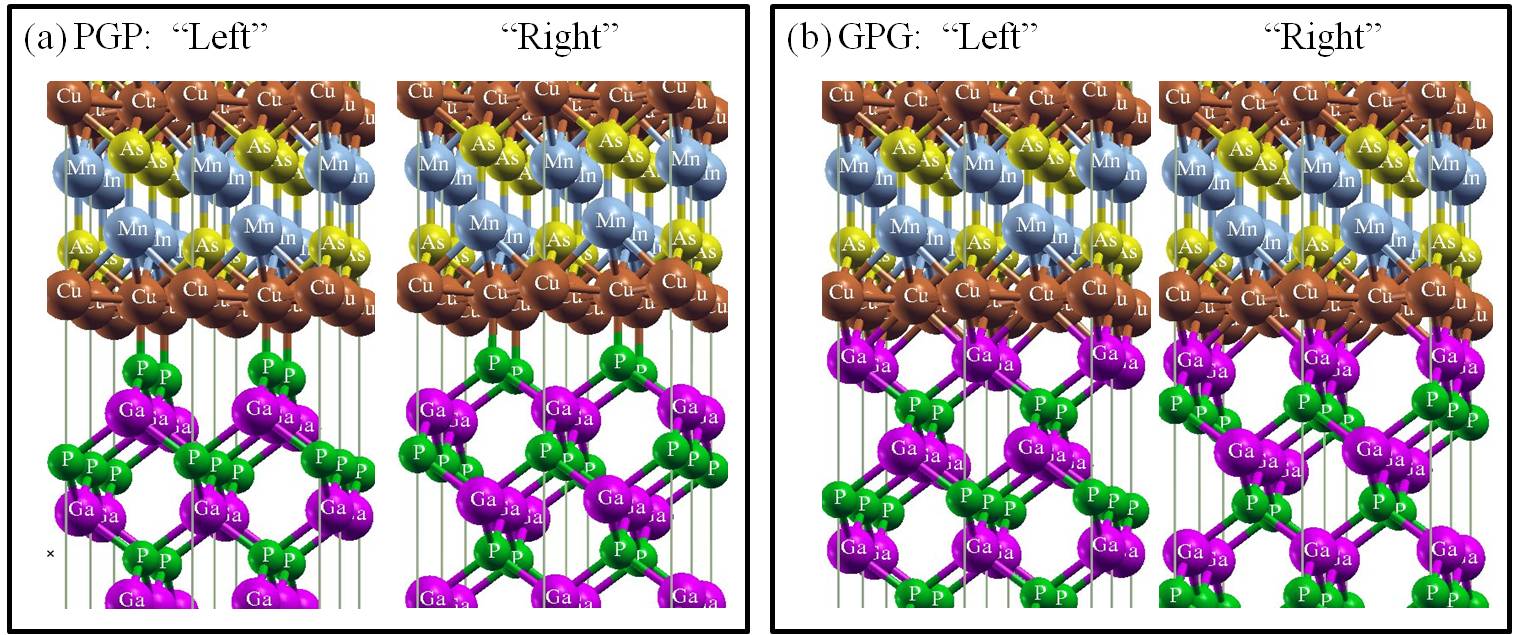}{(Color online) A close view at the left and right-hand-side interfaces of the (a) PGP and (b) GPG junctions 
from the same position in the $x-y$ plane. }{fig06}

%\newpage
\subsection{k-resolved DOS}
In order to understand the nature of the high-transmission states, in Fig. \ref{fig07} we plot the layer and $\vcr{k}_\perp$ resolved spin-PDOS, 
PDOS$_{\bf{k}}$ defined as
\begin{equation}
\mathrm{PDOS}_{\bf{k}}(E)=\frac{1}{\pi V_\mathrm{BZ}} \sum_n^{N_{\bf{k},\mathrm{open}}} \frac{d\bf{k}}{dE}\,,
\end{equation}
where $V_\mathrm{BZ}$ is the BZ volume.

The Mn atoms in layer 22 can be considered to have bulk-like properties, since they are the furthest away from the interface, and indeed the shape 
of the areas of large PDOS match those where the number of propagating states in Fig. \ref{fig04}(a) is non-zero. As one moves close to the interface, 
the Mn DOS changes significantly, demonstrating that interface states are important in this system. The diamond-like feature around the center of the BZ 
is highly changed for the PGP system as one goes towards the Mn interface layer, and no signature of it is visible inside the GaP. The PDOS$_\mathrm{\bf{k}}$ 
inside the GaP has the shape of the interface Mn PDOS, and there is essentially no visible remanence of the bulk Mn PDOS. This shows that for PGP the 
states propagating into the GaP, which dominate the transmission, are mainly interface states. In contrast, for the GPG configuration the diamond-like feature 
is preserved at the interface Mn layer, and it is also visible inside GaP. This indicates that the bulk-like states are important for the transmission in GPG. 
Since interface states generally show a very strong energy dependence, such strong energy dependence is also found in the transmission and TMR of the 
PGP structure. Bulk-like states generally show a less pronounced energy-dependence, and hence the TMR for GPG stack changes less in an energy range 
around $E_\mathrm{F}$.

\myFig{1}{1}{true}{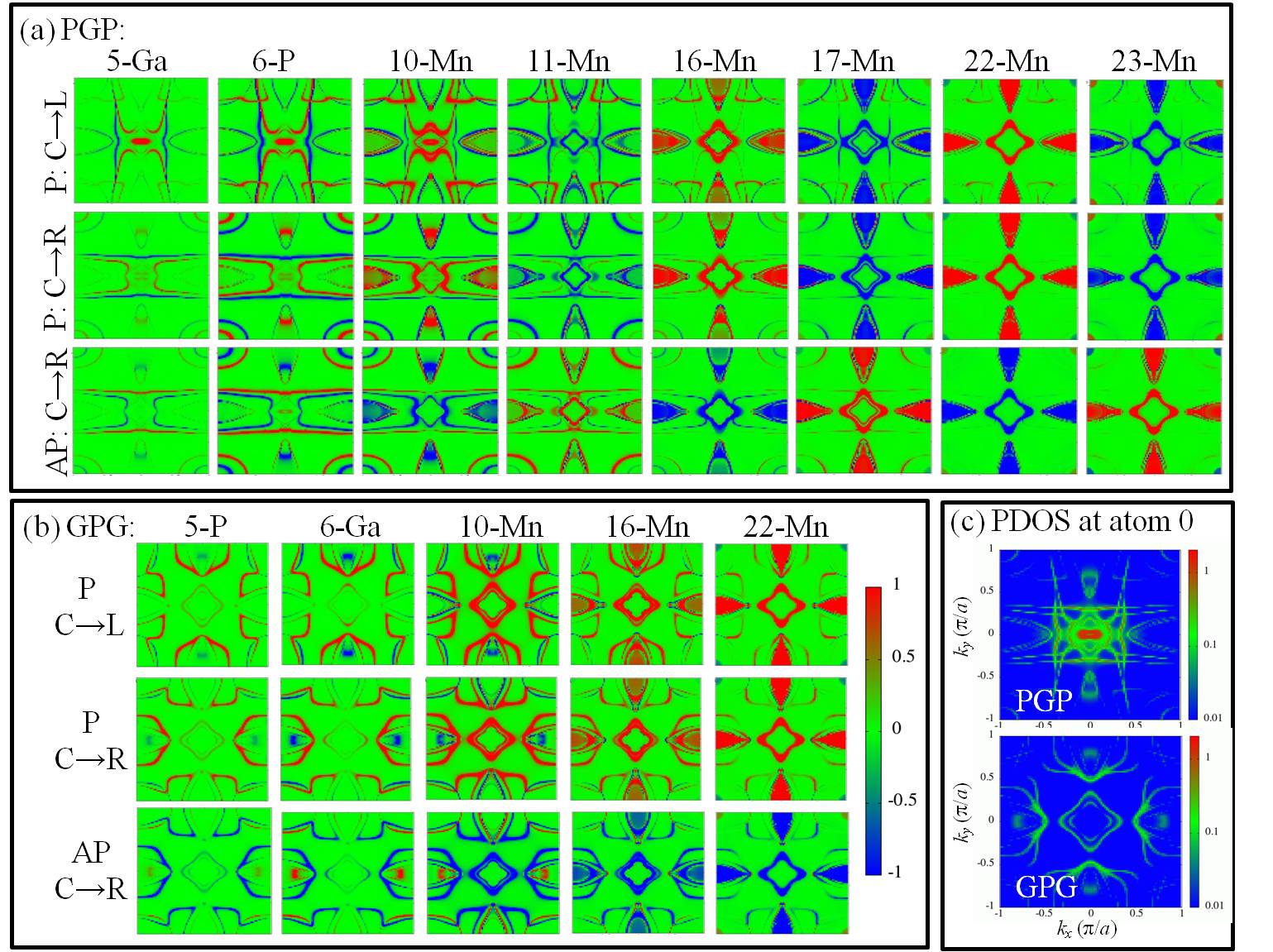}{(Color online) The spin-PDOS over the 2D $(k_x,k_y)$ Brillouin zone at selected atomic sites as indicated by their position number 
counting from the center of the junction along the direction of transport. (a) For the PGP junction, from top to bottom the three rows of panels correspond to the
P state and ``C$ \rightarrow$ R'' (center-to-right sequence of atoms), the P state and ``C$ \rightarrow$ R'', and the AP state and``C$ \rightarrow$ R'', respectively. 
(b) For the GPG junction fewer atoms are depicted (only one of each pair of Mn atoms as the second Mn atom shows analogous but opposite sign contrast). 
(c) PDOS of the center atom in the junction (position ``0'') in the PGP and GPG case (the ``0-P'' and ``0-Ga'', respectively). }{fig07}


\begin{thebibliography}{5}

\bibitem{CoeyBook}J.M.D.~Coey, 
Magnetism and Magnetic Materials, 
{Oxford University Press}, (Oxford, 2009).

\bibitem{RevAFSpin}T.~Jungwirth, X.~Mart\`i, P.~Wadley and J.~Wunderlich, Nature Nanotech. {\bf 11}, 231 (2016).

\bibitem{CMA_Nat} P. Wadley, V. Nov\'ak, R.P. Campion, C. Rinaldi, X. Mart\`i, H. Reichlov\'a, J. \u{Z}elezn\'y, J. Gazquez, 
M.A. Roldan, M. Varela, D. Khalyavin, S. Langridge, D. Kriegner, F. M\'aca, J. Ma\u{s}ek, R. Bertacco, V. Hol\'y, A.W. Rushforth, 
K.W. Edmonds, B.L. Gallagher, C.T. Foxon, J. Wunderlich and T. Jungwirth, Nature Comm. {\bf 4}, 2322 (2013). 

\bibitem{CMA_Science} P. Wadley, B. Howells, J. \u{Z}elezn\'y, C. Andrews, V. Hills, R. P. Campion, V. Nov\'ak, 
K. Olejn\'{i}k, F. Maccherozzi, S. S. Dhesi, S. Y. Martin, T.Wagner, J. Wunderlich, F. Freimuth, Y. Mokrousov, 
J. Kune\u{s}, J. S. Chauhan, M. J. Grzybowski, A. W. Rushforth, K.W. Edmonds, B.L. Gallagher, 
T. Jungwirth, Science {\bf 351}, 587 (2016).

\bibitem{Zelezny} J. \u{Z}elezn\'y, H. Gao, K. V\'yborn\'y, J. Zemen, J. Ma\u{s}ek, Aur\'elien Manchon, 
J. Wunderlich, Jairo Sinova, and T. Jungwirth, Phys. Rev. Lett. {\bf 113}, 157201 (2014).

\bibitem{CMA_SciRep} P. Wadley, V. Hills, M. R. Shahedkhah, K. W. Edmonds, R. P. Campion, 
V. Nov\'ak, B. Ouladdiaf, D. Khalyavin, S. Langridge, V. Saidl, P. Nemec, A. W. Rushforth, 
B.L. Gallagher, S.S. Dhesi, F. Maccherozzi, J. \u{Z}elezn\'y and T. Jungwirth, 
Sci. Rep. {\bf 5}, 17079 (2015).

\bibitem{siesta} J.M.~Soler, E.~Artacho, J.D.~Gale, A.~Garc{\`i}a, J.Junquera, P.~Ordej{\'o}n, and D.~S{\'a}nchez-Portal, 
J. Phys.: Condens Matter \textbf{14}, 2745 (2002).

\bibitem{smeagol1} A. R. Rocha, V. M. Garc{\`i}a-Su\'arez, S. Bailey, C. Lambert, J. Ferrer and S. Sanvito, Phys. Rev. B {\bf 73}, 085414 (2002).

\bibitem{smeagol2} A. R. Rocha, S. Sanvito, Phys. Rev. B \textbf{70}, 094406 (2004).

\bibitem{smeagol3}I.~Rungger and S.~Sanvito, Phys. Rev. B {\bf 78}, 035407 (2008).

\bibitem{HaneyCoCu} P.M.~Haney, D.~Waldron, R.A.~Duine, A.S.~N\~unez, H.~Guo and A.H.~MacDonald, Phys. Rev. B {\bf 76}, 024404 (2007).

\bibitem{HaneyAF} P.M.~Haney, D.~Waldron, R.A.~Duine, A.S.~N\~unez, H.~Guo and A.H.~MacDonald, Phys. Rev. B {\bf 75}, 174428 (2007).

\bibitem{AFmodel} A.S.~N\'u\~nez, R.A. Duine, P. Haney, and A.H. MacDonald, Phys. Rev. B {\bf 73}, 214426 (2006). 

\bibitem{Julliere}M.~Julli\`ere, Phys. Lett. {\bf 54A}, 225 (1975).

\bibitem{Stiles08} C.~Heiliger and M.D.~Stiles, Phys. Rev. Lett. {\bf 100}, 186805 (2008).

%\bibitem{macdonald} P. M. Haney, D. Waldron, R. A. Duine, A. S. Nunez, H. Guo, and A. H. MacDonald, Phys. Rev. B {\bf 76}, 024404 (2007);  C. Heiliger and M.D. Stiles, Phys. Rev. Lett. {\bf 100}, 186805 (2008) .

\end{thebibliography}
\end{document}